\documentclass[10pt,conference]{IEEEtran}
\usepackage{amsmath}
\usepackage{amsfonts}
\usepackage{epsfig}
\usepackage{amssymb}
\usepackage{cite}
\usepackage{hhline}
\usepackage{multirow}
\usepackage{framed}
\usepackage{xcolor}
\usepackage{lipsum}
\usepackage{enumerate}
\usepackage{color}
\usepackage{graphicx}
\usepackage{subfigure}
\usepackage{hyperref}
\usepackage{algorithmic}
\usepackage[ruled]{algorithm}
\usepackage{booktabs}
\usepackage{bm}

\newcommand{\vect}[1]{\bm{#1}}

\begin{document}

\title{Beam-searching and Transmission Scheduling in Millimeter Wave Communications}
\author{\IEEEauthorblockN{Hossein Shokri-Ghadikolaei, Lazaros Gkatzikis, and Carlo Fischione}
\IEEEauthorblockA{Automatic Control Department, Electrical Engineering and ACCESS \\
KTH Royal Institute of Technology, 10044, Stockholm, Sweden \\
{emails: \{hshokri, lazarosg, carlofi\}@kth.se}}}

%\author{Authors % \thanks{Draft, this version is created on {\today}.}}

% ===========================================================================
% File Name: commands.tex
% File Creation Date: 15.6.1995
% Author: Markku Juntti
% Description: Includes useful commands for Latex edition.
% ===========================================================================

% Theorem-like environments
\newtheorem{defin}{Definition}%[Section]
\newtheorem{theorem}{Theorem}
\newtheorem{prop}{Proposition}
\newtheorem{lemma}{Lemma}%[chapter]
\newtheorem{alg}{Algorithm}%[chapter]
\newtheorem{remark}{Remark}
\newtheorem{example}{Example}
\newtheorem{notations}{Notations}
\newtheorem{assumption}{Assumption}

% Equations, arrays, symbols etc.
\newcommand{\be}{\begin{equation}}
\newcommand{\ee}{\end{equation}}
\newcommand{\ba}{\begin{array}}
\newcommand{\ea}{\end{array}}
\newcommand{\bea}{\begin{eqnarray}}
\newcommand{\eea}{\end{eqnarray}}
\newcommand{\combin}[2]{\ensuremath{ \left( \ba{c} #1 \\ #2 \ea \right) }}
   % number of combinations, i.e., polynomial expansion factor
\newcommand{\diag}{{\mbox{diag}}}
\newcommand{\rank}{{\mbox{rank}}}
\newcommand{\dom}{{\mbox{dom{\color{white!100!black}.}}}}
\newcommand{\range}{{\mbox{range{\color{white!100!black}.}}}}
\newcommand{\image}{{\mbox{image{\color{white!100!black}.}}}}
\newcommand{\herm}{^{\mbox{\scriptsize H}}}  % Hermitian transpose
\newcommand{\sherm}{^{\mbox{\tiny H}}}       % Hermitian for subscripts
\newcommand{\tran}{^{\mbox{\scriptsize T}}}  % transpose
\newcommand{\tranIn}{^{\mbox{-\scriptsize T}}}  % transpose
\newcommand{\card}{{\mbox{\textbf{card}}}}
\newcommand{\asign}{{\mbox{$\colon\hspace{-2mm}=\hspace{1mm}$}}}
\newcommand{\ssum}[1]{\mathop{ \textstyle{\sum}}_{#1}}

% Sets of numbers
\newcommand{\vbar}{\raisebox{.17ex}{\rule{.04em}{1.35ex}}}
\newcommand{\vbarind}{\raisebox{.01ex}{\rule{.04em}{1.1ex}}}
\newcommand{\D}{\ifmmode {\rm I}\hspace{-.2em}{\rm D} \else ${\rm I}\hspace{-.2em}{\rm D}$ \fi}
\newcommand{\T}{\ifmmode {\rm I}\hspace{-.2em}{\rm T} \else ${\rm I}\hspace{-.2em}{\rm T}$ \fi}
\newcommand{\B}{\ifmmode {\rm I}\hspace{-.2em}{\rm B} \else \mbox{${\rm I}\hspace{-.2em}{\rm B}$} \fi}
\newcommand{\Hil}{\ifmmode {\rm I}\hspace{-.2em}{\rm H} \else \mbox{${\rm I}\hspace{-.2em}{\rm H}$} \fi}
\newcommand{\C}{\ifmmode \hspace{.2em}\vbar\hspace{-.31em}{\rm C} \else \mbox{$\hspace{.2em}\vbar\hspace{-.31em}{\rm C}$} \fi}
\newcommand{\Cind}{\ifmmode \hspace{.2em}\vbarind\hspace{-.25em}{\rm C} \else \mbox{$\hspace{.2em}\vbarind\hspace{-.25em}{\rm C}$} \fi}
\newcommand{\Q}{\ifmmode \hspace{.2em}\vbar\hspace{-.31em}{\rm Q} \else \mbox{$\hspace{.2em}\vbar\hspace{-.31em}{\rm Q}$} \fi}
\newcommand{\Z}{\ifmmode {\rm Z}\hspace{-.28em}{\rm Z} \else ${\rm Z}\hspace{-.38em}{\rm Z}$ \fi}

% Functions
\newcommand{\sgn}{\mbox {sgn}}
\newcommand{\var}{\mbox {var}}
\newcommand{\E}{\mbox {E}}
\newcommand{\cov}{\mbox {cov}}
\renewcommand{\Re}{\mbox {Re}}
\renewcommand{\Im}{\mbox {Im}}
\newcommand{\cum}{\mbox {cum}}

% Bold-face Greek letters
\renewcommand{\vec}[1]{{\bf{#1}}}     %equation styles %vector/matrix
\newcommand{\vecsc}[1]{\mbox {\boldmath \scriptsize $#1$}}     %equation styles %vector/matrix
\newcommand{\itvec}[1]{\mbox {\boldmath $#1$}}
\newcommand{\itvecsc}[1]{\mbox {\boldmath $\scriptstyle #1$}}
\newcommand{\gvec}[1]{\mbox{\boldmath $#1$}}

\newcommand{\balpha}{\mbox {\boldmath $\alpha$}}
\newcommand{\bbeta}{\mbox {\boldmath $\beta$}}
\newcommand{\bgamma}{\mbox {\boldmath $\gamma$}}
\newcommand{\bdelta}{\mbox {\boldmath $\delta$}}
\newcommand{\bepsilon}{\mbox {\boldmath $\epsilon$}}
\newcommand{\bvarepsilon}{\mbox {\boldmath $\varepsilon$}}
\newcommand{\bzeta}{\mbox {\boldmath $\zeta$}}
\newcommand{\boldeta}{\mbox {\boldmath $\eta$}}
\newcommand{\btheta}{\mbox {\boldmath $\theta$}}
\newcommand{\bvartheta}{\mbox {\boldmath $\vartheta$}}
\newcommand{\biota}{\mbox {\boldmath $\iota$}}
\newcommand{\blambda}{\mbox {\boldmath $\lambda$}}
\newcommand{\bmu}{\mbox {\boldmath $\mu$}}
\newcommand{\bnu}{\mbox {\boldmath $\nu$}}
\newcommand{\bxi}{\mbox {\boldmath $\xi$}}
\newcommand{\bpi}{\mbox {\boldmath $\pi$}}
\newcommand{\bvarpi}{\mbox {\boldmath $\varpi$}}
\newcommand{\brho}{\mbox {\boldmath $\rho$}}
\newcommand{\bvarrho}{\mbox {\boldmath $\varrho$}}
\newcommand{\bsigma}{\mbox {\boldmath $\sigma$}}
\newcommand{\bvarsigma}{\mbox {\boldmath $\varsigma$}}
\newcommand{\btau}{\mbox {\boldmath $\tau$}}
\newcommand{\bupsilon}{\mbox {\boldmath $\upsilon$}}
\newcommand{\bphi}{\mbox {\boldmath $\phi$}}
\newcommand{\bvarphi}{\mbox {\boldmath $\varphi$}}
\newcommand{\bchi}{\mbox {\boldmath $\chi$}}
\newcommand{\bpsi}{\mbox {\boldmath $\psi$}}
\newcommand{\bomega}{\mbox {\boldmath $\omega$}}

% Bold italic Roman letters
\newcommand{\bolda}{\mbox {\boldmath $a$}}
\newcommand{\bb}{\mbox {\boldmath $b$}}
\newcommand{\bc}{\mbox {\boldmath $c$}}
\newcommand{\bd}{\mbox {\boldmath $d$}}
\newcommand{\bolde}{\mbox {\boldmath $e$}}
\newcommand{\boldf}{\mbox {\boldmath $f$}}
\newcommand{\bg}{\mbox {\boldmath $g$}}
\newcommand{\bh}{\mbox {\boldmath $h$}}
\newcommand{\bp}{\mbox {\boldmath $p$}}
\newcommand{\bq}{\mbox {\boldmath $q$}}
\newcommand{\br}{\mbox {\boldmath $r$}}
\newcommand{\bs}{\mbox {\boldmath $s$}}
\newcommand{\bt}{\mbox {\boldmath $t$}}
\newcommand{\bu}{\mbox {\boldmath $u$}}
\newcommand{\bv}{\mbox {\boldmath $v$}}
\newcommand{\bw}{\mbox {\boldmath $w$}}
\newcommand{\bx}{\mbox {\boldmath $x$}}
\newcommand{\by}{\mbox {\boldmath $y$}}
\newcommand{\bz}{\mbox {\boldmath $z$}}

\newenvironment{Ex}
{\begin{adjustwidth}{0.04\linewidth}{0cm}
\begingroup\small
\vspace{-1.0em}
\raisebox{-.2em}{\rule{\linewidth}{0.3pt}}
\begin{example}
}
{
\end{example}
\vspace{-5mm}
\rule{\linewidth}{0.3pt}
\endgroup
\end{adjustwidth}}

%\newenvironment{test}
%{Test begins here: \begingroup \[}
%{\] \endgroup \\ The example has endedd.}

\maketitle

\begin{abstract}
  Millimeter wave (mmW) wireless networks are capable to support multi-gigabit data rates, by using directional communications with narrow beams. However, existing mmW communications standards are hindered by two problems: deafness and single link scheduling. The deafness problem, that is, a misalignment between transmitter and receiver beams, demands a time consuming beam-searching operation, which leads to an alignment-throughput tradeoff. Moreover, the existing mmW standards schedule a single link in each time slot and hence do not fully exploit the potential of mmW communications, where directional communications allow multiple concurrent transmissions. These two problems are addressed in this paper, where a joint beamwidth selection and power allocation problem is formulated by an optimization problem for short range mmW networks with the objective of maximizing effective network throughput. This optimization problem allows establishing the fundamental alignment-throughput tradeoff, however it is computationally complex and requires exact knowledge of network topology, which may not be available in practice. Therefore, two standard-compliant approximation solution algorithms are developed, which rely on underestimation and overestimation of interference. The first one exploits directionality to maximize the reuse of available spectrum and thereby increases the network throughput, while imposing almost no computational complexity. The second one is a more conservative approach that protects all active links from harmful interference, yet enhances the network throughput by 100\% compared to the existing standards. Extensive performance analysis provides useful insights on the directionality level and the number of concurrent transmissions that should be pursued. Interestingly, extremely narrow beams are in general not optimal.
%What is still missing though, is a unifying approach that would efficiently handle the main challenges of mmW communications, namely high attenuation and deafness. Directional transmission and reception along with efficient transmission scheduling could address those challenges. Existing standards adopt the conservative approach of zero interference, that is, no concurrent transmissions, and hence  do not fully exploit the potential of directionality. In this paper, we formulate the problem of joint beamwidth selection and transmission scheduling in a short range mmW network, with the objective of maximizing sumrate performance. We propose efficient protocols to solve this challenging problem, which are also compatible with existing standards. Our extensive performance analysis provides useful insights on the directionality level and the number of concurrent transmissions that should be pursued. Interestingly, extremely narrow beams is in general not optimal.
%deafness and interference management. In this direction, call for novel models that optimize .we propose that efficient scheduling mechanisms can significantly improve the sumrate performance of the system, by scheduling in the same timeslot multiple users that do not cause significant interference to each other.However, the efficient
\end{abstract}

%*As for the abstract, try to rewrite as follows:
%1) some sentences on the background
%2) one or two sentences on what is the problem
%3)  some sentences on how the problem has been solved and what are the
%challenges in the solution procedure/method
%4) what are the specific results that can be obtained by the solution method
%5) what can be concluded more in general


\section{Introduction}\label{sec: introductions}
Millimeter wave (mmW) communications appear as a promising option to meet the ever growing demand for multi-gigabit data rates, at least over short distances. MmW communications refer to the electromagnetic spectrum between 30 and 300~GHz, which corresponds to wavelengths from 10~mm to 1~mm. Small wavelength enables integration of numerous antenna elements in the current size of radio chips, which in turn promises a significant directivity gain. The main characteristics of mmW are directionality, large bandwidth, but also high attenuation~\cite{Rangan2014Millimeter}.

MmW has been considered lately by several standardization bodies as an ideal candidate for short range communications. Specifically, IEEE~802.15.3 task group 3c \cite{802_15_3c} works on the development of high rate wireless personal area networks (WPAN), whereas IEEE~802.11ad task group~\cite{802_11ad} is focused on wireless local area networks (WLAN). In existing standards, one of the network devices is assigned the role of the coordinator, who schedules transmissions in a centralized manner.  In particular, channel access is determined through a hybrid carrier sense multiple access/collision avoidance (CSMA/CA) and time division multiple access (TDMA) scheme. A superframe consists of three phases. A beacon period, a contention access period, where devices compete to register their requests for channel access to the coordinator, and a channel time allocation period, which is further divided into several time slots and each is assigned to a \emph{single} transmitter-receiver pair.
Existing standards do not exploit the full potential of mmW communications. In fact, high data rates are achieved due to the high signal-to-noise ratio (SNR), which is a result of directional communications, and extended bandwidth availability in mmW bands. Pencil beams, however, promises extensive frequency reuse while simplifies interference management~\cite{Xueli4525934,Singh2011Interference}.

In this paper, we suggest that efficient transmission scheduling mechanisms could significantly improve the network throughput (spectral efficiency), by scheduling multiple transmissions in the same time slot, as long as they do not cause harmful interference to each other. The amount of interference caused depends also on the beamwidths that devices operate with. This introduces an alignment-throughput tradeoff. That is, a narrower beamwidth introduces significant searching overhead, since many directions have to be searched, but provides a higher transmission rate due to higher directivity gains, whereas larger beamwidths speed up search process at the expense of loss in the transmission rate. In order to address those problems, we propose a joint formulation of the beamwidth selection and transmission scheduling problems in mmW communications, and analyze the impact of each of the system parameters on the network throughput.

%In addition, we propose that multiple transmissions can be scheduled together, given that
%major challenges of mmW communications, namely deafness and interference management
%A receiver keeps listening to a sector, while transmitter sweeps all possible sectors.
%
%On the positive side, the deafness substantially reduces interference~\cite{Singh2011Interference}, as the receiver only listen to specific directed spatial channel, enabling concurrent transmissions and thus a substantial increased spectral efficiency.

\subsection{Related Work}
Our work is focused on capturing the major tradeoffs in mmW communications, which are mainly due to beam-searching overhead and concurrent transmission scheduling. In the following, we present an overview of existing works in the field.

A main issue in mmW communications is deafness, which is a direct consequence of directional transmission and reception. It occurs when the main beams of a transmitter and the intended receiver are not aligned. To address this issue, beam-searching has been proposed to establish a communication link~\cite{li2013efficient,wang2009beam}. In this case, an exhaustive search over all possible combinations of transmission and reception directions is performed  through a sequence of pilot transmissions.
In fact, mmW devices adopt analog beamforming, also called beam-searching, using simple phase shifters, rather than a complex digital beamforming based on instantaneous channel state information, since the latter would impose formidable complexity in mmW due to the large number of antennas~\cite{Rangan2014Millimeter}. Although the beam-searching concept facilitates the beamforming phase, it introduces an alignment overhead, that is, the time required for finding the best beams.
This overhead depends on the number of directions that have to be searched, which in turn depends on the selected transmission and reception beamwidths.
Current standardization activities~\cite{802_15_3c,802_11ad} suggest a two-stage beam-search technique, to reduce alignment overhead and power consumption. Initially, a coarse grained sector-level sweep is performed, followed by a beam-level alignment phase. An exhaustive search over all possible transmission and reception directions is applied in each level.
For a given beamwidth (fixed granularity of searching),~\cite{li2013efficient} suggests a new search technique as a replacement of the two-stage exhaustive search to reduce the alignment overhead.
Here, we suggest that the alignment-throughput tradeoff should be addressed by optimizing beamwidth per se, thus our work and~\cite{li2013efficient} are complementary.

%given that it consumes a non-negligible portion of the time available for transmission. Regarding scheduling, in order to avoid interference, a single transmission is scheduled within each time slot, which results in highly suboptimal spectral efficiency.

%To the best of our knowledge, the problem of deciding the optimal training period has been only considered for spatial multiplexing in the context of multiple-input-multiple-output systems~\cite{hassibi2003much}. However, due to increased complexity, as mentioned, the existing standards exploit beam-searching based on the pre-defined beam steering vectors could be a feasible solution~\cite{li2013efficient}. Beam-searching is realized through a sequence of pilot transmission between two communicating devices, and hence it introduces significant alignment overhead.
%This overhead depends on the number of directions that have to be searched, that is, by the selected transmission and reception beamwidths. Thus, a portion of the available time for transmission is consumed by beam-searching. In contrast to~\cite{li2013efficient}, where a new search approach is proposed for a given beamwidth, here we optimize beamwidth per se by modeling its impact on effective communication rate. Thus, our work and~\cite{li2013efficient} are complementary.

The option of activating concurrent transmissions for optimally exploiting the directionality of mmW communications was proposed only recently. The authors of~\cite{qiao2012stdma} consider the problem of maximizing the number of scheduled flows such that their quality of service requirement is not violated. A greedy scheduling scheme is proposed, where in each time slot an additional link is activated if its contribution to total throughput is positive, that is, throughput gain from this additional link is larger than the interference caused. A similar greedy heuristic is proposed in~\cite{wu2013flashlinq}, where a priority ordering of links is assumed. Additional links are activated according to this priority order and as long as signal-to-interference-plus-noise ratio (SINR) at all receivers exceeds a threshold. The main issue of all those approaches is that they are reactive protocols, that is, a link has to be activated to deduce if it is compatible with other transmissions. Instead, here we demonstrate that directionality and high attenuation in mmW communications can be exploited to derive accurate scheduling mechanisms.

%  Since a portion of the available time for transmission is consumed by this beam searching procedure, which consequently affects the effective communication rate.
%
%In mmW communications though, the selected beamwidth determines also the amount of time required until a transmitter and a receiver are aligned. Alignment is achieved through a beam searching procedure, that is,
%explicitly address the impact of beamwidth selection on actual performance.

% \cite{qiao2014MAC} has proposed a heuristic algorithm with similar idea. New communication links can be established concurrently, if they do not destroy the ongoing links.
% The main problem with these protocols is a new link should receive feedback from all active receivers to check their SINR values. While this huge overhead is tolerable in omnidirectional control channel within a single hop topology, directivity requirements of mmW networks makes those solution infeasible. A transmitter cannot receive feedback from all active receiver in order to revise its action. In other words, although the existing solution may perform well in other settings, huge signaling costs, which is due to deafness problem, necessitates the reduction of the required signaling and demands new solution for concurrent transmission management.
%
%The proper protocol design for mmW communications requires rethinking of traditional medium access layer (MAC) layer design.

\subsection{Our Contribution}
The main contributions of this paper are summarized into the following
\begin{itemize}
\item We identify the tradeoffs and the corresponding controls that differentiate mmW communications from other technologies.
\item We provide a unifying optimization-based framework that brings together beam-searching and transmission scheduling and explicitly addresses the major challenges of mmW communications, namely deafness and interference management.
\item We demonstrate how the proposed framework can be translated into protocols that extend the capabilities of existing standards.
\item We evaluate the performance gains arising from the proposed protocols. Our performance analysis provides useful insights on the directionality level and the number of concurrent transmissions that should be supported.
\end{itemize}

%----------------------------figure-------------------------------
%\begin{figure}[t]
%	\centering
%	\subfigure[]{
%		\includegraphics[width=7.5cm]{TimingStructureA}
%		\label{subfig: TimingStructureA}
%	}
%	\subfigure[]{
%		\includegraphics[width=7.5cm]{TimingStructureB}
%		\label{subfig: TimingStructureB}
%	}	
%  \caption{Network time segmentations: \subref{subfig: TimingStructureA} IEEE~802.15.3c superframe structure, and \subref{subfig: TimingStructureB} IEEE~802.11ad beacon interval structure.}
%  \label{fig: TimingStructure}
%\end{figure}
%-----------------------------------------------------------------

\section{System Model and Problem Formulation}\label{sec: system model}
\begin{figure}[]
\centering
  \includegraphics[width= 7.5cm]{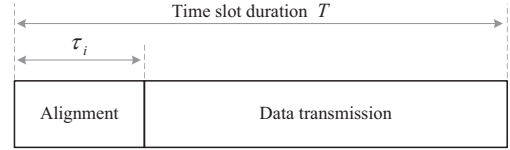}\\
  \caption{Time slot segmentation of each link $i$. Increasing alignment time $\tau_i$ reduces time available for data transmission, but reduces the deafness problem.}
  \label{fig: TimeSlotStructure}
\end{figure}

Consider an mmW network consisting of one coordinator and $N$ transmitter-receiver pairs (links). As shown in Fig.~\ref{fig: TimeSlotStructure}, a time slot consists of two phases: $i)$ alignment and $ii)$ data transmission. Without loss of generality, we assume that sector-level alignment has been established prior to alignment phase, i.e., as part of routing~\cite{singh2009blockage}. In the first phase, transmitter and receiver of each link $i$ have to decide on the optimal refined beams within their sectors, by searching over all possible combinations. This exhaustive search is compliant with IEEE~802.15.3c where all possible combinations within a sector are tested to find the alignment that maximizes SNR.

\subsection{Modelling Alignment Overhead}
Let $T_p$ denote the time required for a pilot transmission, which has to be performed for every combination, and $\psi_{i}^{t}$, $\varphi_{i}^{t}$, $\psi_{i}^{r}$, and $\varphi_{i}^{r}$ be sector-level and beam-level beamwidths at the transmitter and receiver sides of link $i$, respectively. Therefore, under exhaustive search, the total duration of this searching (alignment) procedure is
\begin{equation}\label{eq: search-time}
\tau_i \left( \varphi_{i}^{t}, \varphi_{i}^{r} \right)=  \left\lceil\frac{\psi_{i}^{t}}{\varphi_{i}^{t}}\right\rceil \left\lceil\frac{\psi_{i}^{r}}{\varphi_{i}^{r}} \right\rceil T_p \:,
\end{equation}
where $\lceil \cdot \rceil$ is the ceiling function, returning the smallest following integer. Notice that the number of pilots has to be integer. Once the optimal directions for transmission and reception have been determined, the communication link can be established, and data transmission phase starts. We assume that after the alignment procedure, the transmitter and receiver always find a path to establish data communications, even through a reflection if the direct link is not available~\cite{pyo2009throughput}.
By discarding noncontinuous ceiling function, we derive a continuous approximation of $\tau_i$. The latter cannot exceed total time slot duration $T$, and hence a lower bound on feasible beamwidths can be derived, namely
\begin{equation*}\label{eq: tau-min}
{\varphi_{i}^{t}}{\varphi_{i}^{r}} \geq  \frac{T_p}{T} {\psi_{i}^{t}} {\psi_{i}^{r}} \:.
\end{equation*}
%\LG{this constraint can be discarded and let the optimization neglect those negative solutions. However, the ceiling function has to be discarded later to derive the product form optimality.}
Besides, since alignment takes place within the sector-level beamwidths, we have ${\varphi_{i}^{t}} \leq {\psi_{i}^{t}} $ and ${\varphi_{i}^{r}} \leq {\psi_{i}^{r}}$.

%We assume that the channel quality remains unchanged during the slot, thus once it has been estimated, we can use that estimation for entire slot.
%----------------------------figure-------------------------------
%-----------------------------------------------------------------
%----------------------------figure-------------------------------
\begin{figure}[]
\centering
  \includegraphics[width= 7.5cm]{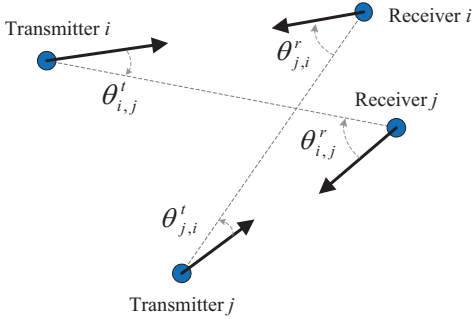}\\
  \caption{Illustration of the angles between transmitters and receivers $\theta_{i,j}^{t}$ and $\theta_{i,j}^{r}$. Solid arrows correspond to the boresight angle.}
  \label{fig: TXRX}
\end{figure}
%-----------------------------------------------------------------
%------------------------------table1-----------------------------
\begin{table}
  \centering
  \caption{Summary of main notations}\label{table: notations}
{
\renewcommand{\arraystretch}{1.2}
  %\scalebox{0.95}
  {
   \begin{tabular}{|l| l| l| l|}
\hline
   \textbf{Symbol} & \textbf{Definition} \\ \hline
%    $i$ & Link index \\ \hline
    $T$ & Time slot duration \\ \hline
    $T_p$ & Pilot transmission time \\ \hline
    $N$ & Number of links \\ \hline
    ${\rm{SINR}}_{i}$ & Signal-to-interference-plus-noise ratio of link $i$\\ \hline
    \multicolumn{1}{|l|}{\multirow{2}{*}{$\theta_{i,j}^{t}$}}  & \multicolumn{1}{|l|}{Angle between transmitter $i$ and receiver $j$}  \\
    \multicolumn{1}{|l|}{} & \multicolumn{1}{|l|}{relative to transmitter boresight angle} \\ \hline
    \multicolumn{1}{|l|}{\multirow{2}{*}{$\theta_{i,j}^{r}$}}  & \multicolumn{1}{|l|}{Angle between receiver $j$ and transmitter $i$}  \\
    \multicolumn{1}{|l|}{} & \multicolumn{1}{|l|}{relative to receiver boresight angle} \\ \hline
    $\varphi_{i}^{t}$ & Beam-level transmitter beamwidth of link $i$ \\ \hline
    $\varphi_{i}^{r}$ & Beam-level receiver beamwidth of link $i$ \\ \hline
    $\psi_{i}^{t}$ & Sector-level transmitter beamwidth of link $i$ \\ \hline
    $\psi_{i}^{r}$ & Sector-level receiver beamwidth of link $i$ \\ \hline
    $p_{i}$ & Transmission power of transmitter $i$ \\ \hline
    $p^{\max}$ & Maximum transmission power \\ \hline
    %$r_{i,j}$ & Received power in receiver $j$ from transmitter $i$  \\ \hline
    $n$ & Noise power \\ \hline
%    $f_{i}$ & Achievable bit rate (bit/sec.) in link $i$ with ${\rm{SINR}}_{i}$ \\ \hline
    $\tau_{i}$ & Channel establishment delay \\ \hline
    $g_{i}^{t}$ & Antenna gain at transmitter $i$ \\ \hline
    $g_{j}^{r}$ & Antenna gain at receiver $j$ \\ \hline
    $g_{i,j}^{c}$ & Channel gain between transmitter $i$ and receiver $j$ \\ \hline
%    $\alpha$ & Path loss index \\ \hline
%    $h_{i,j}$ & Nakagami fading coefficient between transmitter $i$ and receiver $j$ \\ \hline
%    $l_{i,j}$ & Link length between transmitter $i$ and receiver $j$\\ \hline
\end{tabular}}
}
\end{table}

\subsection{Modelling Effective Transmission Rate}
Let $g_{i,j}^{c}$ denote channel gain between transmitter of link $i$ and receiver of link $j$ (in short, transmitter $i$ and receiver $j$), capturing both path loss and block fading, $n$  the power of white gaussian noise, and $p_i$ the transmission power of transmitter $i$. Table~\ref{table: notations} summarizes the main notations used throughout the paper.
For analytical tractability, we approximate the actual antenna patterns by a sectored antenna model, which is a common assumption~\cite{hunter2008transmission,wildman2014joint}.
This simple model captures directivity gains, the front-to-back ratio, and the halfpower beamwidth, which are considered the most important features of an antenna pattern. In ideal sector antenna pattern, the gains are a constant for all angles in the main lobe, and equal to a smaller constant in the side lobe. Let $\theta_{i,j}^{t}$ and $\theta_{i,j}^{r}$ be the angles between transmitter $i$ and receiver $j$ relative to their respective boresight angle (see Fig.~\ref{fig: TXRX}). We denote by $g_{i,j}^{t}$ and $g_{i,j}^{r}$ the transmission and reception gains at transmitter $i$ and receiver $j$ toward each other, that is
\begin{equation}\label{eq: antenna_gain_tx}
g_{i,j}^{t} \left(\theta_{i,j}^{t}, \varphi_{i}^{t} \right) = \left\{ {\begin{array}{*{20}{l}}
\frac{2 \pi - (2\pi - \varphi_{i}^{t})z}{\varphi_{i}^{t}} \:, & {\text{if}~{|\theta_{i,j}^{t}|} \le \frac{\varphi_{i}^{t}}{2}}\\
z \:, &{\text{otherwise}}
\end{array}} \right. \:,
\end{equation}
and
\begin{equation}\label{eq: antenna_gain_rx}
g_{i,j}^{r} \left(\theta_{i,j}^{r}, \varphi_{i}^{r} \right) = \left\{ {\begin{array}{*{20}{l}}
\frac{2 \pi - (2\pi - \varphi_{j}^{r})z}{\varphi_{j}^{r}} \:, & {\text{if}~{|\theta_{i,j}^{r}|} \le \frac{\varphi_{j}^{r}}{2}}\\
z \:, &{\text{otherwise}}
\end{array}} \right. \:,
\end{equation}
where $0 \leq z < 1$ is the gain in the side lobe, with $z \ll 1$ for narrow beams. The gain in the main lobe can be derived by fixing the total radiated power of the antennas over parameter space of $z$, $\varphi_{i}^{t}$, and $\varphi_{i}^{r}$. Then, the received power at receiver $j$ from transmitter $i$ is $p_{i} g_{i,j}^{t} g_{i,j}^{c} g_{i,j}^{r}$, which depends on $p_{i}$, $\varphi_{i}^{t}$, and $\varphi_{j}^{r}$.
%Then, the received power at receiver $j$ from transmitter $i$ is
%\begin{equation*}
%r_{i,j} \left( p_{i},\varphi_{i}^{t}, \varphi_{j}^{r} \right) = p_{i} g_{i,j}^{t} g_{i,j}^{c} g_{i,j}^{r} \:,
%\end{equation*}
Signal to Interference and Noise Ratio (SINR) at receiver of link $i$ is %(\vect{\varphi}^{t}, \vect{\varphi}^{r},\vect{p})
\begin{equation}\label{eq: SINR}
{\rm SINR}_{i} = \frac{p_{i} g_{i,j}^{t} g_{i,j}^{c} g_{i,j}^{r}} {\sum\limits_{\begin{subarray}{c}
  k=1 \\
  k \neq i
\end{subarray}}^{N} {p_{k} g_{k,i}^{t} g_{k,i}^{c} g_{k,i}^{r} + n}}  \:.
\end{equation}
We assume that interference can be treated as noise at each receiver $i$, implying that, according to Shannon formula, link $i$ can achieve a rate of $\log_2 \left( 1 + {\rm SINR}_{i} \right)$ for the remaining $T-\tau_i$ seconds, which can be normalized by time slot duration $T$, to derive the normalized throughput within a time slot.
%\begin{equation}\label{eq: throughput}
%{\text{Throughput}} &= \left(1 - \frac{\tau_i}{T} \right)\log_2 \left( 1 + {\rm SINR}_{i} \right) \:.
%\end{equation}

%\item Considering slotted structure of existing standards,  assign each subsequent mini-slots to several transmission links that do not make harmful interference to each other, in a proactive manner.
%  \item From another perspective, increasing the beamwidth of the main-beam $\theta$ speeds up search process and link establishment, thus throughput enhancement, at the expense of lower directivity gain and higher interference with other devices, thus throughput reduction. Two extreme cases are (1) pencil beam, which wastes almost all transmission time for search process of link establishment, and (2) omnidirectional transmission, which increases interference and avoids concurrent transmissions, but reduces link set-up overhead. The optimal $\theta$ is something in between.

Equation \eqref{eq: SINR} indicates that narrower transmission and reception beamwidths lead to higher directivity gain and hence a higher data rate. As dictated by \eqref{eq: search-time}, this gain is obtained at the expense of higher alignment time $\tau_i$ and consequently less time for data transmission. This arises a tradeoff between the time devoted to alignment phase and the effective data rate. Notice also that decisions of different links are coupled through SINR, and hence scheduling multiple parallel transmissions within a time slot is non-trivial.

\subsection{Maximizing Network Throughput}
In this paper, we consider the problem of joint beamwidth selection and transmission scheduling that has to be solved by the coordinator in each time slot. In particular, we consider a generalized version of the latter where the optimal transmission power of each transmitter has to be selected such that the network throughput (or equivalently spectral efficiency) is maximized. If we collect all control variables $\varphi_{i}^{t}$, $\varphi_{i}^{r}$, and $p_{i}$ in vectors $\vect{\varphi}^{t}$, $\vect{\varphi}^{r}$, and $\vect{p}$, respectively, the problem under consideration can be formally stated as
%, which is determined by transmission/reception beamwidths, and the effective data rate throughput.
%The tradeoff becomes more complex when we include transmission powers $p_i$ in the picture. From traditional communication networks, we know that higher transmission power increases the transmission rate of individual user to some extend, however it introduces extra interference to other links, leading to a complicated power control problem~[\textbf{cite a paper on power control}].
%\begin{align} \label{eq: optimization-multiple-link-oracle}
%& \underset{{\vect{\varphi}^{t}, \vect{\varphi}^{r}, \vect{p}}}{\text{maximize}} \hspace{0.55cm} R=\sum\limits_{i=1}^{N}{{\left( 1 - \frac{\tau_{i}}{T}\right)}{\log_2 \left( 1 + {\rm SINR}_{i} \right)}} \\
%& {\text{subject to}} \hspace{0.6cm} \varphi_{i}^{t} \leq \psi_{i}^{t} \:, \hspace{1.8cm} { 1 \leq i \leq N } \:, \tag{5.a} \\
%& \hspace{2cm} \varphi_{i}^{r} \leq \psi_{i}^{r} \:, \hspace{1.75cm} { 1 \leq i \leq N } \:, \tag{5.b} \\
%& \hspace{2cm} \psi_{i}^{t} \psi_{j}^{r}  {T_P}/{T} \leq \varphi_{i}^{t}\varphi_{j}^{r}  \:, \hspace{0.17cm} { 1 \leq i,j \leq N } \:, \tag{5.c} \\
%& \hspace{2cm} 0 \leq p_{i} \leq p^{\max} \:, \hspace{0.88cm} { 1 \leq i \leq N }  \tag{5.d} \:.
%\end{align}
\begin{subequations} \label{eq: optimization-multiple-link-oracle}
\begin{equation}
\underset{{\vect{\varphi}^{t}, \vect{\varphi}^{r}, \vect{p}}}{\text{maximize}} \hspace{0.55cm} R=\sum\limits_{i=1}^{N}{{\left( 1 - \frac{\tau_{i}}{T}\right)}{\log_2 \left( 1 + {\rm SINR}_{i} \right)}}
\end{equation}
\begin{equation}
{\text{subject to}} \hspace{0.6cm} \varphi_{i}^{t} \leq \psi_{i}^{t} \:, \hspace{1.8cm} { 1 \leq i \leq N } \:,
\end{equation}
\begin{equation}
\hspace{2cm} \varphi_{i}^{r} \leq \psi_{i}^{r} \:, \hspace{1.75cm} { 1 \leq i \leq N } \:,
\end{equation}
\begin{equation}
\hspace{2.4cm} \psi_{i}^{t} \psi_{j}^{r}  {T_P}/{T} \leq \varphi_{i}^{t}\varphi_{j}^{r}  \:, \hspace{0.17cm} { 1 \leq i,j \leq N } \:,
\end{equation}
\begin{equation}
\hspace{2.1cm} 0 \leq p_{i} \leq p^{\max} \:, \hspace{0.88cm} { 1 \leq i \leq N } \:.
\end{equation}
\end{subequations}
Notice that for notational simplicity, function arguments have been discarded.
%{\color{blue}Although outage is not explicitly addressed in~\eqref{eq: optimization-multiple-link-oracle}, it can be easily captured through a modified non-continuous transmission rate function~\cite{chathu}. However, the tradeoffs identified and characterized in this paper would still arise.}
Antenna beamwidths affect both $\tau_i$ and ${\rm{SINR}}_{i}$, whereas transmission powers only affect the latter. Optimization problem formulated in~\eqref{eq: optimization-multiple-link-oracle} is generally non-convex, and hence solving it is a challenging task. In addition, ${\rm{SINR}}_{i}$ and consequently the objective function depend on the network topology, as dictated by $\theta_{i,j}^{t}$ and $\theta_{i,j}^{r}$ in \eqref{eq: antenna_gain_tx} and~\eqref{eq: antenna_gain_rx}. Such information cannot be available at the coordinator in most of WPAN and WLAN systems. %The precise localization problem becomes more challenging for mobile devices operating in an indoor environment, a typical practical scenario.
In the next section, we investigate structural properties of problem~\eqref{eq: optimization-multiple-link-oracle}, which enable us to propose two standard-compliant and easy to implement algorithms of low complexity.

\section{Joint Beamwidth Selection and Transmission Scheduling}\label{sec: optimization}
The optimization problem formulated in~\eqref{eq: optimization-multiple-link-oracle} is generally non-convex,
making it difficult to be optimally solved. To derive some insight on the arising tradeoffs, we first focus on the single link case ($N=1$), which is also the case for existing mmW standards~\cite{802_15_3c}. Next, we consider the general problem of concurrent transmissions and we demonstrate that it can be reduced to multiple parallel single link instances.%, followed by complexity analysis of optimizing the above tradeoffs.

%Thereby, we get some insights about the objective function and extract its critical features. Besides, the existing standards activate only one link per time slot, which is similar to our single link scenario.

\subsection{Single Link Scenario}\label{subsec: single-link}
Consider a network consisting of a single link $i$, where no interference is experienced by the receiver. For a given time slot this is also the case of existing standards. Once alignment procedure has been completed, both transmitter and receiver operate in their main lobes, hence $\theta_{i,i}^{t} = \theta_{i,i}^{r} =0$. This implies that SINR expression, formulated in~\eqref{eq: SINR}, reduces to
\begin{equation}\label{eq: SINR-single-link}
{\rm SNR}_{i} = \frac{g_{i,i}^{c}}{n} p_{i} \left(\frac{2 \pi - (2\pi - \varphi_{i}^{t})z}{\varphi_{i}^{t}} \right) \left( \frac{2 \pi - (2\pi - \varphi_{i}^{r})z}{\varphi_{i}^{r}} \right) \:.
\end{equation}
%Notice that SNR is independent of network topology, since $\theta_{i,i}^{t} = \theta_{i,i}^{r} =0$. Thus, problem ~\eqref{eq: optimization-multiple-link-oracle} is significantly simplified.

\emph{\textbf{Proposition I:}} Consider problem~\eqref{eq: optimization-multiple-link-oracle} for a single link scenario, where $N=1$ and SINR is replaced by SNR according to equation~\eqref{eq: SINR-single-link}. Optimal transmission power in this case is $p^{\max}$.

\emph{Proof:} Since increasing transmission power does not affect the alignment overhead, yet monotonically enhances ${\rm SNR}_{i}$, transmitting at the maximum power is optimal $p_i^*=p^{\max}$ . \hspace*{\fill}{$\blacksquare$}

\emph{\textbf{Result I:}} Consider optimization problem~\eqref{eq: optimization-multiple-link-oracle} for single link scenario. For parameters in the region of interest, the optimal antenna beamwidths $\left( \varphi_{i}^{t} \right)^*$ and $\left( \varphi_{i}^{r} \right)^*$ can be accurately approximated by a hyperbola $\left( \varphi_{i}^{t}  \right)^* \left( \varphi_{i}^{r}  \right)^* = \varphi_{i}^{*}$, where $\varphi_{i}^{*}$ is a constant that depends on system parameters $\psi_i^t$, $\psi_i^r$, $T_p$, $T$, $p^{\max}$, $z$, $g_{i,i}^{c}$ and $n$.

\emph{Proof:} The proof is provided in Appendix~A. \hspace*{\fill}{$\blacksquare$}

The above results imply that the dimension of the optimization problem in the single link scenario can be reduced from $3$ variables, namely $p_i$, $\varphi_{i}^{t}$ and $\varphi_{i}^{r}$, into a single one, namely $\varphi_{i} \triangleq \varphi_{i}^{t}\varphi_{i}^{r}$. Next, we derive an additional property of the objective function, which validates the existence of a tradeoff between alignment overhead and achievable throughput.

\emph{\textbf{Proposition II:}} Consider optimization problem \eqref{eq: optimization-multiple-link-oracle} for single link scenario. For system parameters in the region of interest, the optimal antenna beamwidth $\varphi_{i}^{*}$ is the unique solution of ${\partial R} / {\partial \varphi_i}=0$  .

\emph{Proof:} The proof is provided in Appendix~B. \hspace*{\fill}{$\blacksquare$}

Proposition II implies that adopting extremely narrow beams is not always optimal in terms of throughput due to the huge alignment overhead. Also, wide beams devastate the directivity gains, and hence they do not provide the maximum throughput. Given channel gain $g_{i,i}^{c}$, the network coordinator can find the optimal beamwidths $\varphi_{i}^{*}$ through a simple gradient descent algorithm~\cite{convex-boyd}. In the next subsection, we show how the coordinator can obtain an estimation of the channel gain between the transmitter and receiver $i$.
We now turn our attention to the multiple links case, and show the steps required to convert a multiple links scenario to single link scenario, so to efficiently address the general optimization problem~\eqref{eq: optimization-multiple-link-oracle}.

\subsection{Multiple Links Scenario}\label{subsec: multiple-links}
Although current standards schedule a single link within each time slot, narrow beams promise significant throughput gain by exploiting concurrent transmissions.
Optimization problem~\eqref{eq: optimization-multiple-link-oracle} provides the maximum network throughput, under the assumption that the coordinator knows the exact topology of the network. To tackle this issue, we propose two topology-agnostic approaches. The first one is a conservative approach that generally overestimates the interference experienced by each link. In the second approach, we perform transmission scheduling by assuming that resulting interference will be negligible, which is supported by the pseudo-wired abstraction of mmW communications~\cite{Singh2011Interference}.

%and find the maximum throughput gain of concurrent transmission.

%The best solution depends on applications and quality of service guarantees of the links.

%Thus, the coordinator avoids any interference in the network at the expense of reducing the maximum achievable throughput, still it is substantially higher than single link assignment, as verified with numerical results.

%\emph{Remark}: Proposition II holds for multiple links case.
%
%\emph{Proof}: The proof is provided in Appendix~A. \hspace*{\fill}{$\blacksquare$}

\vspace{3mm}
\noindent \textbf{Overestimation of interference}
\vspace{1mm}

The main idea behind this approach is to estimate interference at sector level, which is generally higher than interference experienced at beam-level.
An IEEE~802.15.3c or 802.11ad compliant device has to be equipped with an orthogonal frequency division multiplexing (OFDM) transceiver, which enables different links to operate in different frequency channels at the same time. Inspired by FlashLinQ protocol proposed in~\cite{wu2013flashlinq}, we can derive the following low-overhead protocol to estimate interference.
%Let ${\rm SNR}_i$ and ${\rm INR}_{ji}$ denote the signal-to-noise ratio of link $i$ and the interference-to-noise ratio of transmitter $j$ at receiver $i$, respectively.
First, the coordinator assigns orthogonal channels to different links, one to each link. Each transmitter $i$ transmits with power $p^{\max}$ inside its sector and in its dedicated channel, without introducing any interference to other links. Here, we assume that the sector of the intended receiver/transmitter can be derived from a local table~\cite{singh2009blockage}. Each receiver $i$ measures SNR in link $i$, denoted by ${\rm SNR}_i$, and also overhears the received power from every transmitter $j$ with sector-level beam. The latter provides an estimation of the amount of interference-to-noise-ratio from transmitter $j$, denoted by ${\rm INR}_{ji}$.
% These channels can be realized either with different frequencies\footnote{A standard compliant mmW device, either IEEE~802.15.3c or 802.11ad, has OFDM transceiver. Therefore, a coordinator can ask different links to operate in different frequency channels at the same time, without any extra hardware complexity.} or codes.
Then, each link checks if it can be activated with other links concurrently without receiving/making harmful interference. From the analysis provided in~\cite{etkin2008gaussian}, the sufficient condition for link $i$ to be independent of link $j$ (treating interference as noise) is
\begin{equation}\label{eq: sufficient-condition-IS}
\sqrt{{\rm SNR}_i} \geq {\rm INR}_{j,i}, \quad {\text{and}} \quad \sqrt{{\rm SNR}_i} \geq {\rm INR}_{i,j} \:.
\end{equation}
Each receiver $i$ evaluates the interference level from transmitter $j$ and according to sufficient conditions~\eqref{eq: sufficient-condition-IS} creates the set of interferers, that is, the set of links with which link $i$ should not be activated simultaneously. Notice that interference has been estimated at sector level, whereas actual transmissions take place over fine-grained beams. Thus, this is an overestimation of the actual interference during communications with pencil beams, providing a conservative approach that overestimates interference, so to ensure that no collisions occur. The receivers feedback their interferer sets to the coordinator. The coordinator then derives a conflict graph that shows the links that cannot be activated concurrently. Next, we provide a detailed description of the proposed scheme.

A conflict graph $\mathcal{G} = (\mathcal{V}, \mathcal{E})$ is defined by a set of vertices $\mathcal{V}$ and edges $\mathcal{E}$. A vertex $i \in \mathcal{V}$ represents communication link $i$ and an edge $(i,j) \in \mathcal{E}$ indicates that links $i$ and $j$ cannot be activated simultaneously due to high mutual interference. In fact, the interferer set of link $i$, which is reported to the coordinator by receiver $i$, represents the set of neighbors of vertex $i$ in the conflict graph. Finally, an independent set of graph $\mathcal{G}$ is a subset of $\mathcal{V}$ that contains no adjacent vertices, indicating that those vertices (links) can be concurrently activated without any harmful interference. This enables a transformation of power allocation subproblem of ~\eqref{eq: optimization-multiple-link-oracle} to transmission scheduling instance. That is, the coordinator, out of all independent sets, should activate at maximum power the links of the independent set that achieves maximum throughput. For a given independent set and due to mutual independency of its links, the coordinator can optimize each link individually using a simple gradient descent, as already discussed in the single link scenario.

Let $\mathcal{I}_k$ be independent set $k$, and $\mathcal{I}$ be the set of all independent sets. Then, problem~\eqref{eq: optimization-multiple-link-oracle} can be cast as a scheduling problem that activates the links of the independent set that maximizes the network throughput. That is,
%maximizes the throughput of the network~\cite{tassiulas1992stability}, which is our objective in this paper.
%\begin{align} \label{eq: optimization-multiple-link-overestim}
%& \underset{\mathcal{I}_k \subseteq \mathcal{I}, {\mathbf{\varphi^{t}}, \mathbf{\varphi^{r}}}}{\text{maximize}} \hspace{0.45cm} \sum _{i \in \mathcal{I}_k}{{\left( 1 - \frac{\tau_{i}}{T}\right)}{\log_2 \left( 1 + {\rm SINR}_{i} \right)}} \\
%& {\text{subject to}} \hspace{0.6cm} \varphi_{i}^{t} \leq \psi_{i}^{t} \:, \hspace{1.8cm} { 1 \leq i \leq N } \nonumber \\
%& \hspace{2cm} \varphi_{i}^{r} \leq \psi_{i}^{r} \:, \hspace{1.75cm} { 1 \leq i \leq N } \nonumber \\
%& \hspace{2cm} \psi_{i}^{t} \psi_{j}^{r}  {T_P}/{T} \leq \varphi_{i}^{t}\varphi_{j}^{r}  \:, \hspace{0.17cm} { 1 \leq i,j \leq N } \nonumber \:,
%\end{align}
\begin{subequations}\label{eq: optimization-multiple-link-overestim}
\begin{equation}
\underset{\mathcal{I}_k \subseteq \mathcal{I}, {\mathbf{\varphi^{t}}, \mathbf{\varphi^{r}}}}{\text{maximize}} \hspace{0.45cm} \sum _{i \in \mathcal{I}_k}{{\left( 1 - \frac{\tau_{i}}{T}\right)}{\log_2 \left( 1 + {\rm SNR}_{i} \right)}} \hspace{0.6cm}
\end{equation}
\begin{equation}
{\text{subject to}} \hspace{0.6cm} \varphi_{i}^{t} \leq \psi_{i}^{t} \:, \hspace{1.8cm} { 1 \leq i \leq N } \:,
\end{equation}
\begin{equation}
\hspace{2cm} \varphi_{i}^{r} \leq \psi_{i}^{r} \:, \hspace{1.75cm} { 1 \leq i \leq N } \:,
\end{equation}
\begin{equation}
\hspace{2.4cm} \psi_{i}^{t} \psi_{j}^{r}  {T_P}/{T} \leq \varphi_{i}^{t}\varphi_{j}^{r}  \:, \hspace{0.17cm} { 1 \leq i,j \leq N } \:,
\end{equation}
\end{subequations}
where ${\rm SNR}_{i}$ is given by \eqref{eq: SINR-single-link}, since there is no interference inside an independent set.

Given the independent sets, problem \eqref{eq: optimization-multiple-link-overestim} can be solved efficiently
by using gradient descent algorithms. However, finding all independent sets is an NP-hard problem in general~\cite{sharma2006complexity}. For sparse conflict graphs, which is the case in mmW networks with pencil beams, efficient solutions exist~\cite{lozin2008polynomial}. \ref{protocol: protocol-1} describes the required steps to convert the joint beamwidth selection and power allocation problem~\eqref{eq: optimization-multiple-link-oracle} to a joint beamwidth selection and transmission scheduling problem~\eqref{eq: optimization-multiple-link-overestim}.

{
\renewcommand{\baselinestretch}{1.05}
\begin{algorithm} [t]
\floatname{algorithm}{}
\renewcommand{\thealgorithm}{Protocol I}
\caption{\small Interference-aware scheduling in mmW communications}
\label{protocol: protocol-1}
\begin{algorithmic}[1]
\small
\STATE Initially, the coordinator assigns orthogonal channels to different links. A single channel is assigned to each.
\STATE Each transmitter $i$ transmits with power $p^{\max}$ with sector-level beam in its dedicated channel.
\STATE Each receiver $i$ measures received power from transmitter $i$ with sector-level beam and computes ${\rm SNR}_i$.
\STATE Each receiver $i$ overhears the received power from each transmitter $j$ with sector-level beam and computes ${\rm INR}_{j,i}$.
\STATE Each receiver $i$ evaluates sufficient conditions~\eqref{eq: sufficient-condition-IS} and creates the set of interferers.
\STATE All receivers feedback their interferer sets to the coordinator.
\STATE The coordinator creates a conservative conflict graph, and schedules links based on~\eqref{eq: optimization-multiple-link-overestim}.
\end{algorithmic}
\end{algorithm}
}

\vspace{3mm}
\noindent \textbf{Underestimation of interference}
\vspace{1mm}

 According to the pseudo-wired abstraction of mmW communications~\cite{Singh2011Interference}, that is, the condition that a relatively small number of active links operating with narrow beams do not cause interference to each other, we may neglect interference. Thus, we may optimize each link individually, as if it was operating on its own. Then, the problem of joint optimization of antenna beamwidth and transmission power for $N$ links can be decomposed to $N$ parallel single link problems. Each can be solved in polynomial time as described in Section~\ref{subsec: single-link}. \ref{protocol: protocol-2} describes the steps of the proposed underestimation approach.  For an ultra dense network, where interference is non-negligible even under directional communications, an adaptive algorithm could be investigated to modify channel access action of every link upon every failure, but this is beyond the scope of this work.

%%%%%%%%%%%%%%%%%%%%%%%%%%%%%%%%%%%%%%%%%%%%%%%%%%%%%%%%%%%%%%%%%%%%%%%%%%%%%%%%%%%%%
%%%%%%%%%%%%%%%%%%%%%%%%%%%%%%%%%%%%%%%%%%%%%%%%%%%%%%%%%%%%%%%%%%%%%%%%%%%%%%%%%%%%%%%%%%%%%%%%%%%%%%%%%%%%%
{
\renewcommand{\baselinestretch}{1.05}
\begin{algorithm} [t]
\floatname{algorithm}{}
\renewcommand{\thealgorithm}{Protocol II}
\caption{\small Interference-agnostic scheduling in mmW communications}
\label{protocol: protocol-2}
\begin{algorithmic}[1]
\small
\STATE Initially, the coordinator assigns orthogonal channels to different links. A single channel is assigned to each.
\STATE Each transmitter $i$ transmits with power $p^{\max}$ with sector-level beam in its dedicated channel.
\STATE Each receiver $i$ estimates channel gain of link $i$, that is, $g_{i,i}^{c}$.
\STATE All receivers feedback their channel gains.
\STATE The coordinator optimizes beamwidth of every link individually, and each transceiver adjusts its beamwidth accordingly. All transmissions take place at maximum power $p^{\max}$.
\end{algorithmic}
\end{algorithm}
}
%%%%%%%%%%%%%%%%%%%%%%%%%%%%%%%%%%%%%%%%%%%%%%%%%%%%%%%%%%%%%%%%%%%%%%%%%%%%%%%%%%%%%%%%%%%%%%%%%%%%%%%%%%%%%%%
%%%%%%%%%%%%%%%%%%%%%%%%%%%%%%%%%%%%%%%%%%%%%%%%%%%%%%%%%%%%%%%%%%%%%%%%%%%%%%%%%%%%%%%%%%%%%%%%%%%%%%%%%%

\section{Numerical Results}\label{sec: numerical-results}
In order to derive some insight on the behaviour of the identified tradeoffs, we consider scenario of several mmW devices, randomly located in an area of $10 \times 10 ~m^2$, operating in 60GHz with maximum power of $2.5$ mW, which are typical values in bluetooth-based WPAN. We assume $90^{\circ}$ sector-level beams both at transmitter and receiver side. Using Monte Carlo simulations, we evaluate the network throughput, in bits per time slot per hertz, over $100$ random topologies.

%----------------------------figure-------------------------------
\begin{figure}[]
\centering
  \includegraphics[width= 6.5cm]{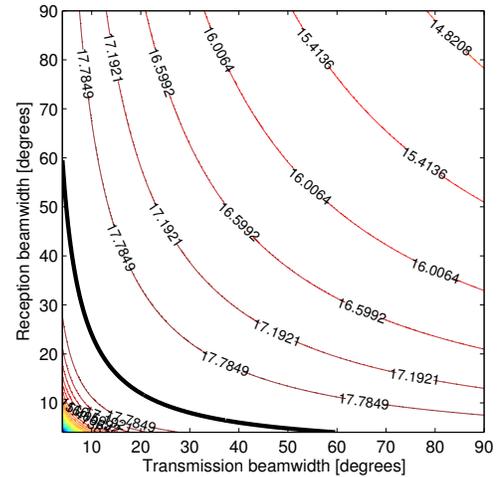}\\
  \caption{Optimal region of transmission and reception beamwidths.}
  \label{fig: Fig1_symmetry}
\end{figure}
%-----------------------------------------------------------------
Fig.~\ref{fig: Fig1_symmetry} illustrates Result I by depicting contours of the throughput of a single link against transmission and reception beamwidths. The bold black curve shows the optimal beamwidth region for which throughput is maximized. This corresponds to $\varphi_{i}^{t}\varphi_{i}^{r} = 240$, for the example considered.
Based on this result, for the rest of simulations, we assume that $\varphi_{i}^{t} = \varphi_{i}^{r}$ for all $i$.

%----------------------------figure-------------------------------
\begin{figure}[]
\centering
  \includegraphics[width= 8.5cm]{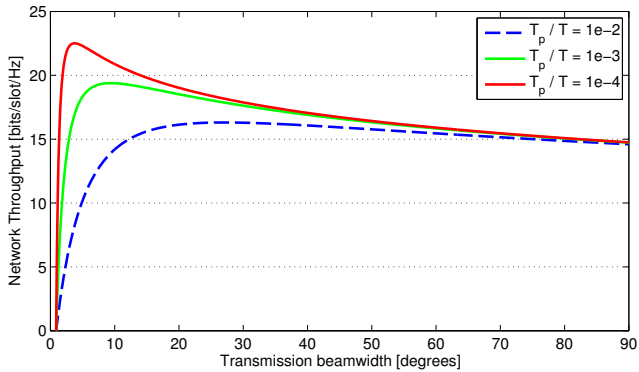}\\
  \caption{Alignment-throughput tradeoff in mmW networks.}
  \label{fig: Fig2-Throughput_phi}
\end{figure}
%-----------------------------------------------------------------

Fig.~\ref{fig: Fig2-Throughput_phi} demonstrates the alignment-throughput tradeoff for a single link mmW network. For narrow beamwidths, beam-searching overhead is dominating, whereas as operating beamwidths increase, directivity gain becomes more important. Generally, the optimal point is a balance between directivity gain over the benefit of additional transmission time.
Moreover, reduced pilot transmission overhead $T_p/T$, enables the transmitter and receiver to execute more beam-searching iterations using the same time budget. As a result, performance is improved, and narrower beams are more beneficial. It is noteworthy that Proposition II holds in all three depicted scenarios. That is, they have positive derivative at the beginning of their feasible set and negative one at the end.

%----------------------------figure-------------------------------
\begin{figure}[]
\centering
  \includegraphics[width= 8.5cm]{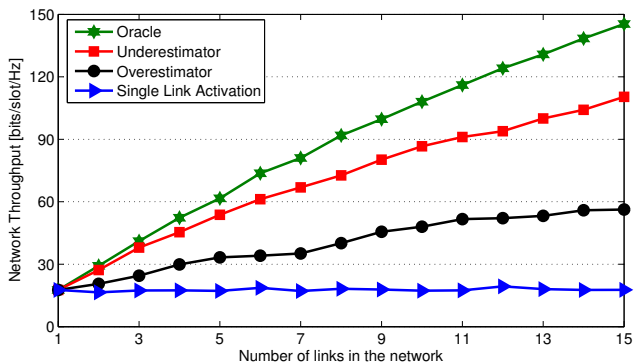}\\
  \caption{Comparison of different concurrent transmission schemes in multiple links scenarios.}
  \label{fig: Fig3_Comparison_multiple_links}
\end{figure}
%-----------------------------------------------------------------
Fig.~\ref{fig: Fig3_Comparison_multiple_links} compares the performance of the proposed schemes for multiple links scenarios. For comparison purposes, we depict \emph{Oracle} which is the solution of optimization problem~\eqref{eq: optimization-multiple-link-oracle}, whereas \emph{Single Link Activation} represents the network throughput achieved if the link of highest SNR is activated. The following points can be made from this figure.
First, allocating only one channel per time slot, which is the case in the existing standards, does not exploit fully the time slots in mmW networks. This inefficiency increases with the number of links. In particular, with $10$ links, 525\%, 401\%, and 177\% performance enhancement can be achieved by the Oracle, interference under-estimator, and over-estimator, respectively. %The throughput gain will be increased by the number of potential links, which demands further analysis on capacity scaling in mmW networks. Second, \emph{Oracle} solution requires unaffordable computational complexity.
Given that the number of links in local networks is limited, typically less than 20 links, the underestimation approach can provide low complexity solutions that are close to the optimal. Good performance is expected for small scale networks, where users operate with narrow beams. For ultra dense networks, however, high amount of interference invalidates the basic pseudo-wired assumption, based on which the proposed underestimation approach has been developed. %For such networks, a combination of SDMA and OFDMA can be a proper multiple access solution.
Instead the conservative approach guarantees that no harmful interference arises at the cost of a significant throughput reduction, yet it outperforms the current single link activation scheme. This gain increases also with the number of links, since a higher number of links increases the probability of having more independent links. In general, deciding which is the most appropriate scheme depends heavily on the computation complexity that can be handled, number of links, and quality of service requirements of individual links.

\section{Conclusion}\label{sec: concluding-remarks}
Millimeterwave (mmW) communications promise a significant improvement in spectral efficiency of next generation wireless networks, if the many challenges arising from the special physical characteristics of mmW frequency bands. In this paper, we demonstrated that taking full advantage of this huge potential requires optimizing alignment-throughput tradeoff and novel solutions for optimal scheduling of concurrent transmissions. The problem of joint beamwidth selection and power control was formulated. It was shown that such a problem cannot be efficiently solved, since the network topology needs to be known.
%with the objective of maximizing spectral efficiency. Solving this optimization problem required full knowledge of the network topology along with exponentially growing computational overheads.
To tackle this, two approaches were proposed based on overestimation and underestimation of interference. Overestimation approach activates only a small number of non-interfering links to ensure that no harmful interference will be received by any active links. This overprotection leads to reduced performance gain, yet doubles the network throughput compared to the existing standards. The underestimation approach by neglecting interference, which is a valid assumption for moderately large mmW networks, yields a close to the optimal performance with light computational complexity.

In this work we proposed standard-compliant algorithms for short range mmW scenarios. Most of the identified tradeoffs arise also in cellular mmW networks, where a hybrid digital-analog beamforming has to be conducted due to formidable complexity of pure digital beamforming design for large number of antennas. Extending the proposed schemes and addressing the additional challenges that arise in the context of cellular networks is an interesting topic for future study.
%Although this paper focuses on increasing spectral efficiency, the framework can be extended to address energy efficiency in mmW communications.

\section*{Appendix A}\label{appen: appendix-A}
To prove Result I, we show that beamwidths appear in the optimization problem only in the form of ${\varphi_{i}^{t} \varphi_{i}^{r}}$. Thus, instead of optimizing over individual $\varphi_{i}^{t}$ and $\varphi_{i}^{r}$, we can optimize the throughput over a new decision variable $\varphi_{i} \triangleq {\varphi_{i}^{t} \varphi_{i}^{r}}$. The above claim holds on alignment procedure, since it can be fully described by $\varphi_{i}$. Considering optimal transmission power $p^{\max}$, we recast~\eqref{eq: SINR-single-link} to
\begin{align*}
{\rm SINR}_{i} &= \frac{g_{i,i}^{c}}{n} p^{\max} \left(\frac{2 \pi - 2\pi z}{\varphi_{i}^{t}} + z \right) \left(\frac{2 \pi - 2\pi z}{\varphi_{i}^{r}} + z \right) \\
               &= c_2 \left(\underbrace{\frac{\left(2 \pi - 2\pi z\right)^2}{\varphi_{i}}}_{S_1} + \underbrace{\frac{\left(2 \pi - 2\pi z \right) z \left( \varphi_{i}^{t} + \varphi_{i}^{r} \right)}{\varphi_{i}}}_{S_2} + z^2 \right)\:,
\end{align*}
where $c_2 = {g_{i,i}^{c}} p^{\max}/n$. Now, if $S_1 \gg S_2$, we can neglect the term $S_2$, so SINR and consequently achievable rate will be completely described only by $\varphi_{i}$. After some algebraic manipulations, we can show that the condition holds if
\begin{equation*}
 \varphi_{i}^{t} + \varphi_{i}^{r} \ll 2 \pi \frac{1-z}{z} \:.
\end{equation*}
For narrow beams, $z \ll 1$, which along with $\varphi_{i}^{t} \leq \psi_{i}^{t}$ and $\varphi_{i}^{r} \leq \psi_{i}^{r}$, confirms the above condition. For example, under realistic assumptions of $\psi_{i}^{t} = \psi_{i}^{r} = \pi/2$ and $z=0.1$, the condition translates to
\begin{equation*}
\varphi_{i}^{t} + \varphi_{i}^{r} \leq \frac{\pi}{2} + \frac{\pi}{2}  \ll 18 \pi \:,
\end{equation*}
which is true. Therefore, we can conclude that for the parameters in the region of interest, term $S_2$ is negligible compared to $S_1$, thus the throughput of link $i$ can be well approximated by
\begin{equation}\label{eq: SINR-single-link-simplified-2}
\left( 1 - \frac{c_1}{\varphi_{i}}\right)\log_2 \left( 1 + c_2 z^2 + c_2 \frac{\left(2 \pi - 2\pi z\right)^2}{\varphi_{i}} \right) \:,
\end{equation}
where $c_1 = \psi_{i}^{t}\psi_{i}^{r}T_p /T$ and $c_2$ are two constants within a time slot.

It can be easily shown that Result I holds also for the multiple links scenario. For each transmitter $k$ and receiver $i$ that do not suffer from deafness, received interference can be accurately characterized with the product $\varphi_{k}^{t}\varphi_{i}^{r}$ using the same claims as for the single link case. If another transmitter $n$ and receiver $i$ are not correctly aligned (deafness scenario), thus at least one of them produces gain $z \ll 1$, we can neglect the contribution of the interference from transmitter $n$ on the aggregated interference of the receiver $i$ if $z$ is small enough. \hspace*{\fill}{$\blacksquare$}

\section*{Appendix B}\label{appen: appendix-B}
In order to prove Proposition II, we show that the derivative of the throughput has up to one root, i.e.,  the throughout has up to one extremum.
Derivative of~\eqref{eq: SINR-single-link-simplified-2} with respect to $\varphi_i$ is
\begin{align}\label{eq: drivative-expression}
\frac{\partial\eqref{eq: SINR-single-link-simplified-2}}{\partial \varphi_i} = \hspace{1mm}& \frac{c_1}{\varphi_{i}^2} \log_2 \left( 1 + c_2 z^2 + c_2 \frac{\left(2 \pi - 2 \pi z\right)^2}{\varphi_{i}} \right) \\
& - \frac{ \left( 1 - \frac{c_1}{\varphi_i} \right) c_2 \left(2 \pi - 2 \pi z\right)^2 }{ {\varphi_{i}^2} \ln(2)\left( 1 + c_2 z^2 + c_2 \frac{\left(2 \pi - 2 \pi z\right)^2}{\varphi_{i}} \right)} \nonumber \:,
\end{align}
where $c_1$ and $c_2$ are as defined in Appendix A. Given that $\varphi_i > 0$ and after some algebraic manipulations, we derive that the roots of the derivative are the solutions of the following equation
\begin{align}\label{eq: roots-of-drivative}
&\frac{c_1}{\varphi_{i}^2} \log_2 \left( 1 + c_2 z^2 + c_2 \frac{\left(2 \pi - 2 \pi z\right)^2}{\varphi_{i}} \right) \\
&\hspace{5mm}= \frac{c_2 \left(2 \pi - 2 \pi z\right)^2 }{\ln(2)} \frac{\varphi_i - c_1}{ \left(1 + c_2 z^2 \right) \varphi_i + c_2 \left(2 \pi - 2 \pi z\right)^2} \nonumber \:.
\end{align}
Whereas the left hand side of~\eqref{eq: roots-of-drivative} is a strictly decreasing function of $\varphi_i$, it is easy to show that the right hand side is a strictly increasing function. Therefore, right and left hand sides has up to one crossing point $\varphi_i^*$, which is the extremum of the throughput, formulated in~\eqref{eq: SINR-single-link-simplified-2}.

Next, we prove that this extremum is a maximum for the parameters in the region of interest.
It is sufficient to show that~\eqref{eq: drivative-expression} takes a positive value at the beginning and a negative value at the end of the feasible set. Therefore, given that we have up to one extremum, there is only one $\varphi_i^*$ inside the feasible set, and not on the boundaries, for which the throughput will be maximized.
From the constraints of the optimization problem~\eqref{eq: optimization-multiple-link-oracle}, the minimum and maximum possible values of $\varphi_i$ are $c_1$ and $c_1T/T_p$. If $\varphi_i \rightarrow c_1^+$, \eqref{eq: drivative-expression} tends to
\begin{equation*}\label{eq: derivative-SINR-single-link}
\frac{1}{c_1} \log_2 \left( 1 + c_2 z^2 + c_2 \frac{\left(2\pi - 2 \pi z\right)^2}{c_{i}} \right) \:,
\end{equation*}
which is strictly positive, given that $\log_2(\cdot)$ represents non-negative data rate. On the other hand, for $\varphi_i$ close to the maximum value $c_1 T/T_p$, we note that the time left for transmission becomes almost $(1-T_p/T)$ that can be well approximated by $1$, since $T$ is the time slot duration which should be much higher than the transmission of a single pilot $T_p$.
Neglecting the alignment overhead $(1-c_1/\varphi_i)$ from~\eqref{eq: SINR-single-link-simplified-2}, the throughput will be a strictly decreasing function of $\varphi_{i}$, confirming that the derivative takes negative values at the end of the feasible set. This completes the proof. \hspace*{\fill}{$\blacksquare$}

\bibliographystyle{IEEEtran}
\bibliography{IEEEabrv,References}

\end{document}